\documentclass[11pt]{article}
\begin{document}
\title {Power-Law Energy Splitting Generated By Tunneling Between
Non-smooth Tori}
\author{Zai-Qiao Bai \\
Institute of Theoretical Physics, Academia Sinica, Beijing 100080, China}

\date{}
\maketitle
\baselineskip 18pt \vskip 5mm
\begin {minipage}{140mm}
\begin {center} {\bf Abstract} \end {center}
\baselineskip 18pt

 We discuss the energy level splitting $\Delta\epsilon$ due to
 quantum tunneling between congruent tori in phase space. In
 analytic cases, it is well known that $\Delta\epsilon$ decays
 faster than power of $\hbar$ in the semi-classical limit. This is
 not true in non-smooth cases, specifically, when the tori are
 connected by line on which the Hamiltonian is not smooth. Under
 the assumption that the non-smoothness depends only upon the x- or
 p-coordinate, the leading term in the semi-classical expansion of
 $\Delta\epsilon$ is derived, which shows that $\Delta \epsilon$
 decays as $\hbar^{k+1}$ when $\hbar\rightarrow 0$ with $k$ being
 the order of non-smoothness.

\end {minipage}
\newpage

\section{Introduction}
 This paper concerns the splitting of classically degenerate energy
 levels. The near degeneracy (ND) classically corresponds to
 congruent tori in phase space while the quantum tunneling
 between the tori causes the splitting \footnote{In this paper, the
 word ``tunneling" refers to the quantum transition between states
 that classically correspond to separate tori in phase space \cite
 {Gut}.}. A well-known example occurs in the one-dimensional
 symmetric double-well potential, where the eigenenergies below the
 top of the barrier cluster into two-fold ND's with energy
 difference vanishes as
 $$
 \Delta \epsilon \sim \hbar^{\alpha}{\rm
 e}^{-S/\hbar} \eqno (1.1)
 $$
 when $\hbar\rightarrow 0$. When
 turning to multi-dimensional cases, M. Wilkinson showed that
 $\Delta \epsilon$ vanishes normally in the same or, in certain
 situation, even more singular manner than (1.1)\cite{Wks1}.
 However, is it always true that the energy level splitting
 resulting from quantum tunneling is smaller than any power of
 $\hbar$ in the semi-classical limit? Let us see the following
 example.

 Consider the system on one-dimensional circle defined by any of
 the four Hamiltonians,
 $$
 H_{1}=\frac{p^2}{2}+\cos^2 x,~~~
 H_{2}=\frac{p^2}{2}+|\cos x|,~~~
 H_{3}=|p|+\cos^2 x,~~~
 H_{4}=|p|+|\cos x|.
 \eqno(1.2)
 $$
 In classical mechanics, the
 above Hamiltonians determine similar phase space portraits,
 particularly, the motion at $H\neq 1$ contains two symmetric
 closed orbits: the two vibrational orbits with $H<1$ are connected
 by the transition $(x,p)\rightarrow (x+\pi,p)$ and the two
 rotational orbits with $H>1$ are connected by the time reversal
 $(x,p)\rightarrow (x,-p)$. According to the
 Einstein-Brillouion-Keller (EBK) semi-classical quantization rule,
 this classical degeneracy implies a two-fold ND structure in the
 spectrum of $H$. We can verify this prediction by directly
 diagonalizing the Hamiltonians. In Fig.1, $\Delta
 \epsilon$ is plotted as the function of the mean energy of the ND
 pair($\epsilon$). As expected, $\Delta \epsilon$ (open dots) is much
 smaller than the spacing of $\epsilon$ (approximately the dotted
 lines). However, contrary to the exponentially decay of $\Delta
 \epsilon$ with the increase of $|\epsilon-1|$ in Fig.1(a),
 non-exponentially decay of $\Delta \epsilon$ in some cases is
 obvious. From the four illustrations, we can see that the
 ``exceptional" ND occurs when and only when the corresponding
 classically degenerate tori (closed orbits) in phase space are
 connected by line(s) where the Hamiltonian is not smooth. This
 fact suggests that tunneling between the degenerate tori can be
 greatly enhanced by the passage of non-smoothness.

  In fact, M.V. Berry showed this non-smoothness-enhanced quantum
  transition between classically degenerate states about two
  decades ago\cite{Berry}. In studying the coefficient $r$ for
  reflection above a barrier $V(x)$ in the semi-classical
  limit, Berry proved that $r\sim \hbar^k$ when $V(x)$ has a
  discontinuous $k$-th derivatives, in contrast to the analytic
  case where $r$ is exponentially small. Another
  interesting quantum manifestation of non-smoothness, the
  power-law localization of eigenstates was also discussed
  in more recent papers(e.g. \cite{Rbm,Jung,Li}).

  In this paper we shall investigate the energy level splitting
  resulting from the non-smoothness-enhanced tunneling. We first
  consider the
  case where ND is related to the time reversal symmetry. By
  perturbation method, a relation between $\Delta \epsilon$ and the
  non-smoothness of the potential is derived. Based on a geometrical
  interpretation, this relation is applied to a class of
  non-smooth systems.

\section {Power-Law Energy Splitting}
  In this section we study systems where ND is related to the
  time reversal symmetry. The problem is more tractable since
  the projection of torus onto the coordinate space contains no
  singularity (caustic). By perturbation method, we obtain
  an explicit power-law $\hbar$-dependence of the energy splitting.

  Consider a mechanic system on one-dimensional circle with
  Hamiltonian $H=E_k(p)+V(x)$, $V(x+2\pi)=V(x)$. The kinetic energy
  $E_k(p)$ satisfies $E_k(-p)=E_k(p)$, and, for simplicity, we
  assume $E_k(0)=0$, $E_k(\infty)=\infty$ and
 $\frac{d}{dp}E_k(p)>0$ when $p>0$. A familiar example of
 such kinetic energy is $\frac{1}{2}p^2$. Due to the time
 reversal symmetry, the two classical orbits at
 $H(x,p)=E>\max_x V(x)$, $O^{+}_{E}$ and $O^{-}_{E}$, one
 with $p>0$ and the other with $p<0$, yield identical action
 integral, i.e.,
 $$
 \oint_{O^{+}_{E}} pdx =\oint_{O^{-}_{E}} pdx=S(E).
 \eqno(2.1)
 $$
 Consequently, EBK quantization
 condition $S(E)=2n\pi\hbar$ predicts a two-fold degenerate
 level $E=\epsilon_n$. The two semi-classical eigenfunctions
 are given by
 $$
 \Psi_{n}^{\pm}(x)=\frac{1}{\sqrt{T_n
 \dot{x}_n}}\exp[\pm is_n(x)/\hbar],
 \eqno(2.2)
 $$
 where
 $s_{n}(x)=\int_0^xp_{n}(x')dx'$, $p_{n}(x)>0$ is determined
 by $E_k(p)+V(x)=\epsilon_n$, $\dot{x}_n=\frac{d}{d p}
 E_{k}(p)|_{p=p_n(x)}$ is the classical velocity and the
 normalization constant
 $T_n=\int_0^{2\pi}\frac{dx}{\dot{x}_n}$ is the period of
 the corresponding classical orbit{\cite{Almd}}. (The suffix
 ``n" of $\epsilon$, $\Psi$, $p$, $\dot{x}$, $T$ and $s$
 will be hereafter dropped out for simplicity.)

 Of course, in general, the two levels do not exactly coincide. The
 difference between $\epsilon$ and the exact eigenenergy is of
 order $o(\hbar)$ in the semi-classical limit ($\hbar\rightarrow
 0,n\rightarrow \infty$ while $n\hbar$ is fixed). In the case that
 $V(x)$ is not smooth (infinitely differentiable), we have seen in
 the last section ($H_2$ and $H_4$) that the splitting of energy
 levels ($\Delta \epsilon$) is not exponentially small. It is
 therefore possible that a non-vanishing $\Delta \epsilon$ will
 emerge from the higher order semi-classical corrections. If we are
 only interested in the leading term in $\Delta \epsilon$, however,
 variational calculation in the space spanned by $\Psi^{+}$ and
 $\Psi^{-}$ will give the result.
 We shall consider a simple case that $V(x)$
 is a $C^{k-1}$ function and
 $$
 {\bigwedge}_x^{k}V(x)
 \equiv \lim_{x^{\prime} \rightarrow x+0} \frac{d^k}{dx^k}
 V(x^{\prime})- \lim_{x^{\prime} \rightarrow x-0} \frac{d^k}{dx^k}
 V(x^{\prime})
 \eqno(2.3)
 $$ is well-defined, which vanishes on
 $[0,2\pi]$ except at discrete points
 $x_j^{\ast},j=1,...,N<\infty$. Then elementary calculations show
 that the energy splitting is given by (see Appendix )
 $$
 \Delta \epsilon=
 \frac{\hbar^{k+1}}{2^{k}T}|\sum_{j=1}^{N}\frac{\exp(2is(x_j^{\ast})/\hbar)}
 { p^{k+1}\frac{d}{d p} E_{k}|_{p=p(x_j^{\ast})}}
 {\bigwedge}_x^{k}V(x_j^{\ast})|+o(\hbar^{k+1}) \equiv \Delta
 \epsilon^{(0)}+o(\hbar^{k+1}).
 \eqno(2.4)
 $$
 Define a dimensionless measurement of ND by $\eta_n=\frac{2\Delta
 \epsilon_n}{\epsilon_{n+1}-\epsilon_{n-1}}$. Noticing that
 the semi-classical level spacing is
 $2\pi\hbar\frac{dE}{dS} =2\pi\hbar/T$ and according to Eq.
 (2.4), we find
 $$
 \eta=\frac{\hbar^{k}}{2^{k+1}\pi}
 |\sum_{j=1}^{N}\frac{\exp(2is(x_j^{\ast})/\hbar)}
 { p^{k+1}\frac{d}{d p} E_{k}|_{p=p(x_j^{\ast})}}
 {\bigwedge}_x^{k}V(x_j^{\ast})|+o(\hbar^{k}) \equiv
 \eta^{(0)}+o(\hbar^{k}).
 \eqno(2.5)
 $$

 {\sl Example 2.1} $ H=\frac{1}{2}p^2+V^{(k)}(x)$, where
 $ V^{(1)}(x)=\max\{\cos x,0\} $
 and $V^{(k)}(x)=[V^{(1)}(x)]^k$, $k=2,3,...$.\\
 According to Eq. (2.5), when $\epsilon>1$,
 $$
 \eta^{(0)}=
 \frac{k!\hbar^{k}}{2^{k}\pi(2\epsilon)^{\frac{k}{2}+1}}
 |\sin[\frac{(2\epsilon)^{\frac{1}{2}} \pi}{\hbar}+\frac{k\pi}{2}]|.
 $$
 The comparison of $\eta$ and $\eta^{(0)}$ is shown in Fig.2.

{\sl Example 2.2} $ H=|p|+V^{(k)}(x)$.\\
 When $\epsilon>1$, the semi-classical
 level is given by $\epsilon_n=n\hbar+\alpha_k$ and according to
 Eq. (2.4)
 $$
 \Delta\epsilon^{(0)}=
 \frac{k!\hbar^{k+1}}{2^k\pi\epsilon^{k+1}}
 |\sin(\frac{\alpha_k\pi}{\hbar}+\frac{k\pi}{2})|
 $$
 where
 $$
 \alpha_k\equiv \frac{1}{2\pi}\int_{-\frac{\pi}{2}}
 ^{\frac{\pi}{2}}\cos^kx dx=\frac{\Gamma(\frac{k+1}{2})}
 {2\Gamma(\frac{1}{2})\Gamma(\frac{k}{2}+1)}.
 $$
 The comparison of $\Delta\epsilon$ and $\Delta\epsilon^{(0)}$
 is shown in Fig.3.

\section {Sum Over Transition Paths}
 In this section we first give Eq. (2.4) a geometrical
 interpretation. We find the quantum transition between the
 semi-classical eigenstates can be classically described by the
 leaking of phase space points from one torus to the other via
 passage of non-smoothness. This picture will facilitate the
 generalization of Eq. (2.4).

 The splitting of nearly degenerate energy levels is closely
 related to the transition probability between the corresponding
 semi-classical eigenstates.
 In classical picture, $\Psi^{+}$ describes a particle
 moving on the circle with $p>0$. After one classical period,
 due to quantum tunneling, the particle has a
 non-zero probability to jump to
 the reflection wave $\Psi^{-}$  with $p<0$.
 Write $\exp (\frac{HT}{i\hbar})|\Psi^{+}>=c|\Psi^{+}>+
 {\cal A}|\Psi^{-}>$. Simple calculations show that
 ${\cal A}\approx \frac{T}{\hbar}<\Psi^{-}|(H-\epsilon)|\Psi^{+}>$
 and $\Delta \epsilon \approx \frac{2\hbar}{T}|{\cal A}|$.
 According to Eq. (A.17), the leading term in ${\cal A}$
 is the sum of contribution from each non-smooth point of
 $V(x)$, i.e.,
 $$
 {\cal A}\approx {\cal A}^{(0)}=
 \sum_{j=1}^{N}r_j\exp(i\phi_j),
 \eqno(3.1) $$
 with
 $$
 r_j=\frac{(i\hbar)^{k}}
 {(2p^{\ast}_j)^{k+1} \dot{x}_j^{\ast}}
 {\bigwedge}_x^{k}V(x_j^{\ast})
 {\rm ~~~and~~~}\phi_j=2s(x_j^{\ast})/\hbar,
  \eqno(3.2)
 $$
 where $p^{\ast}_j\equiv p(x^{\ast}_j)$ and
 $\dot{x}_j^{\ast}\equiv\frac{d}{d p} E_{k}|_{p=p(x_j^{\ast})}$.
 We note that $r_j$ is exactly the reflection coefficient
 obtained by Berry \footnote{Berry's calculation was based on
 $E_k=\frac{p^2}{2}$. However, the result (Eq. (27) in \cite{Berry})
 is essentially identical to Eq. (3.2).}.

 As the classical representation of $\Psi^{+}$ and $\Psi^{-}$, the
 tori $O_{\epsilon}^{+}$ and $O_{\epsilon}^{-}$ are connected by
 the straight line $x=x_j^{\ast}$ where $H$ is not smooth. We shall
 call the vector on $x=x_j^{\ast}$ that starts from
 $O_{\epsilon}^{+}$ and ends at $O_{\epsilon}^{-}$ a {\sl
 transition path} and denote it by $\gamma_j$ (Fig.4). Accordingly,
 we can say that $\Psi^{+} \rightarrow\Psi^{-}$ is dominated by the
 tunneling along transition path(s). In fact, the reflection
 coefficient $r_j$ is determined by the local properties of
 $\gamma_j$. Besides a constant, $r_j$
 consists of three ingredients. ${\bigwedge}_x^{k}V(x_j^{\ast})$
 can be regarded as the intensity of non-smoothness at $\gamma_j$.
 $\frac{1}{(2p^{\ast}_j)^{k+1}}$ describes a power-law
 decay with the increase of path length
 $2p^{\ast}_j=\hbar \frac{\partial \phi_j} {\partial x_j^{\ast}}$.
 $\frac{1}{\dot{x}_j^{\ast}}$, which comes from the product of
 amplitude of semi-classical wave functions, gives a classical
 weight of the transition path: the longer the particle stays in
 the vicinity of the non-smooth point, the more probably it
 jumps to the other torus. In contrast to $r_j$,
 the phase $\phi_j$ is not determined by the local properties
 of $\gamma_j$.
 Since only the relative phase is of physical importance, i.e.,
 gives rise to interference effect, we find
 $$
 \phi_j-\phi_k=\frac{2}{\hbar} (s(x_j^{\ast})-s(x_k^{\ast}))
 =\frac{1}{\hbar}\oint_{\gamma_{jk}}p d x,
 \eqno (3.3)
 $$ where
 $\gamma_{jk}$ is a closed path consists of $\gamma_{j}$,
 -$\gamma_{k}$ ($\gamma_{k}$ with opposite direction) and the
 segments of $O_{\epsilon}^{+}$ and $O_{\epsilon}^{-}$ (real paths)
 that attached at their ends (see Fig.4). If $\gamma_{jk}$ is
 contractible, $\phi_{j}-\phi_{k}$ is simply the phase
 space area (in the unit of $\hbar$) enclosed by this closed path.

 Behind the simple form of Eq. (3.2) there are two non-generic
 facts resulting from the assumption that $\frac{d}{d p}E_k(p)>0$
 when $p>0$: the starting and end points of $\gamma_j$ are
 symmetric with respect to $p=0$ and the projection of
 $O_{\epsilon}^{+}$ or $O_{\epsilon}^{-}$ onto the coordinate space
 contains no singularity. Now we ignore this assumption and require
 only $E_k(-p)=E_k(p)$ to guarantee the time reversal symmetry. Let
 $A_j=(x^{\ast}_j,p_j)\in O_{\epsilon}^{+}$ and
 $A_j^{\prime}=(x^{\ast}_j,p_j^{\prime}) \in O_{\epsilon}^{-}$ be
 the starting and end points of $\gamma_j$. By adopting the general
 semi-classical eigenfunctions corresponding to the tori
 $O_{\epsilon}^{+}$ and $O_{\epsilon}^{-}$ \cite {Almd}, similar
 calculations as that performed in Appendix show that Eq. (3.2-3)
 should be modified as
 $$
 {r_j}=\frac{(i\hbar)^{k}}
 {(p_j-p_j^{\prime})^{k+1} \sqrt{|\dot {x}(A_j)
 \dot{x}(A_j^{\prime})|}} {\bigwedge}_x^{k}V(x_j^{\ast})
 \eqno(3.4)
 $$
 and
 $$
 \phi_j-\phi_k=\frac{1}{\hbar}\oint_{\gamma_{jk}}p d x
 -M_{jk}\pi/2,
 \eqno (3.5)
 $$ where $M_{jk}$ is the sum of the
 Maslov indices of the segments of real paths on $\gamma_{jk}$.
 Having the contribution of each transition path, we need only to
 sum over all these paths to obtain the energy splitting
 $\Delta \epsilon ^{(0)}$ or $\eta^{(0)}$.

 {\sl Example 3.1} $ H=(p^2-1)^2+V(x)$,
 where $V(x)=1-(\frac{x}{\pi})^2, |x|\leq \pi$. \\
 When $\epsilon <1$, the Maslov index of
 $O_\epsilon^{+}$ (or $O_\epsilon^{-}$), which encircles
 point $(\pi,1)$ (or $(\pi,-1)$), is $2$ and
 EBK quantization condition reads
 $S(\epsilon_n)=2(n+\frac{1}{2})\pi \hbar$ (see inset of Fig. 5).
 The straight line $x=\pi$ intersects $O_\epsilon^{+}$
 (or $O_\epsilon^{-}$) at points $A_{1,2}$ (or
 $A_{1,2}^{\prime}$)
 where $p=(1\pm \epsilon^{\frac{1}{2}})^{\frac{1}{2}}$
(or $-(1\pm \epsilon^{\frac{1}{2}})^{\frac{1}{2}}$).
 There exist four transition paths, i.e.,
 $\gamma_1$ ($A_1\rightarrow A_1^{\prime}$),
 $\gamma_2$ ($A_2\rightarrow A_2^{\prime}$),
 $\gamma_3$ ($A_1\rightarrow A_2^{\prime}$) and
 $\gamma_4$ ($A_2\rightarrow A_1^{\prime}$).
 Moreover, EBK quantization condition implies
 $\phi_{j}-\phi_{1}=0,n\pi,n\pi$ (mod $2\pi$) for $j=2,3,4$
 respectively.
 According to Eq. (3.4), up to a phase,
 $$
 {\cal A}^{(0)}=\frac{i\hbar}{4\pi{\epsilon}^{\frac{1}{2}}}
 [\frac{1}{(1+{\epsilon}^{\frac{1}{2}})^{\frac{3}{2}}}
 +\frac{1}{(1-{\epsilon}^{\frac{1}{2}})^{\frac{3}{2}}}
 +(-1)^n\frac{4}{(1+{(1-\epsilon)}^{\frac{1}{2}})
 (1-\epsilon)^{\frac{1}{4}}}].
 $$
 When $\epsilon >1$, only $\gamma_1$ survives
 so that
 $$
 {\cal A}^{(0)}=\frac{i\hbar}{4\pi{\epsilon}^{\frac{1}{2}}
 (1+{\epsilon}^{\frac{1}{2}})^{\frac{3}{2}}}.
 $$
 Numerical results show that $\eta^{(0)}=|{\cal A}^{(0)}|/\pi$
 is a good approximation of $\eta$ when $\hbar$ is sufficiently
 small and $\epsilon$ is not too close to $1$, the energy of
 separatrix(Fig.5).

 The same treatment can be applied to non-smooth systems where ND
 is originated from spatial symmetries. By substituting
 $(p,-x)\rightarrow (x,p)$, relations (3.4-5) can be directly
 transformed to systems where the non-smoothness that results
 transition path depends only upon the p-coordinate. Specifically,
 consider  a transition path $\gamma_j$ on the straight line
 $p=p_{j}^{\ast}$ with starting and end points at
 $A_j=(x_j,p_j^{\ast})$ and
 $A_j^{\prime}=(x^{\prime}_j,p_j^{\ast})$, the corresponding
 reflection coefficient should be
 $$
 r_j=\frac{(i\hbar)^{k}}
 {(x_j^{\prime}-x_j)^{k+1} \sqrt{|\dot {p}(A_j)
 \dot{p}(A_j^{\prime})|}}{\bigwedge}_p^{k}H(x,p_j^{\ast}).
 \eqno(3.6)
 $$
 The phase difference is also given by Eq. (3.5),
 whereas the Maslov index should count the singularity of the
 projection of torus onto the momentum space \footnote{ We use
 $\omega_1=p d x$ instead of $\omega_1^{\prime}=-x d p$ based on
 two facts. Firstly,
 $\oint_{\gamma_{jk}}\omega_1=\oint_{\gamma_{jk}}\omega_1^{\prime}$
 when $\gamma_{jk}$ is contractible. Secondly, if the coordinate
 space has non-trivial topology, $\omega_1$ is well-defined while
 $\omega_1^{\prime}$ is not. We find that this choice is justified
 by numerical results.}. Despite this similarity, interesting
 behavior may occur when the configuration space has a non-trivial
 topology. We shall demonstrate it by some examples.

 Suppose the configuration space is a circle, i.e., $(x,p)$ and
 $(x+2\pi,p)$ describe the same point. In this case,
 a path $(x,p)\rightarrow(x',p)$ implies a family of
 paths $(x,p)\rightarrow(x'+2n\pi,p), n\in Z$.
 If we attribute the contribution of all these paths
 to a representative path, say, $(x,p)\rightarrow(x',p)$,
 the only change of Eq.(3.6) is that
 $\frac{1}{(x_j^{\prime}-x_j)^{k+1}}$
 should be replaced by
 $$
 \sum _{q=-\infty}^{\infty}\frac{\exp(i2q\pi p^{\ast}_j/\hbar)}
 {(x_j^{\prime}-x_j+2q\pi)^{k+1}}
 \equiv
 W_{k+1}(x_j^{\prime}-x_j,p^{\ast}_j/\hbar).
 \eqno(3.7)
 $$
 $W$ satisfies periodic condition
 $W_k(x,y+1)=e^{i2\pi y}W_k(x+2\pi,y)=W_k(x,y)$.
 When $y\in[0,1]$, $$
 \begin {array}{l}
 \displaystyle W_2(x,y)=\frac{1}{4\sin ^2\frac{x}{2}}
 [1+y(e^{ix}-1)]e^{-ixy}, \\\\
 \displaystyle W_3(x,y)=\frac{1}{8\sin ^3\frac{x}{2}}
 [\cos \frac{x}{2}+i2y \sin \frac{x}{2}-2y^2
 \sin^2 \frac{x}{2}e^{ix/2}]e^{-ixy}\\
 \end {array}
 \eqno(3.8)
 $$
 and so on. We note that ${\cal A}^{(0)}$ is in
 general not invariant under the translation $(x,p)\rightarrow
 (x,p+\delta_p)$ when $\delta_p$ is not an integer multiple of
 $\hbar$, which is however always a symmetric transformation in
 classical mechanics. This difference reflects the discreteness of
 quantum momentum space.

 {\sl Example 3.2} $H=|p-p_c|+\cos^{2}x $. \\ The symmetric
 double-well potential causes ND at $\epsilon<1$. According
 to Eq. (3.6-7), the total contribution of transition paths
 (on $p=p_c$) is given by
 $$
 {\cal A}^{(0)}=\frac{i\hbar}{{\epsilon}^{\frac{1}{2}}
 (1-\epsilon)^{\frac{1}{2}}} [W_2(2x_c,\frac{p_c}{\hbar})
 +W_2(2\pi-2x_c,\frac{p_c}{\hbar})
 +(-1)^{n}2W_2(\pi,\frac{p_c}{\hbar})],
 $$ where
 $x_c=\cos^{-1}{\epsilon}^{\frac{1}{2}}$. When $p_c=0$,
 ${\cal A}^{(0)}=\frac{i\hbar}{2{\epsilon}^{\frac{1}{2}}
 {(1-\epsilon)}^{\frac{1}{2}}}
 [\frac{1}{1-\epsilon}+(-1)^{n}]$. When
 $p_c=\frac{\hbar}{2}$, ${\cal A}^{(0)}=0$. In fact, $\Delta
 \epsilon \equiv 0$ in this case because $H$ is represented
 by the same matrix in the invariant subspaces spanned by
 functions $\{e^{i2nx}\}_n$ and
 $\{e^{i(2n+1)x}\}_n$ respectively.

 Consider a spin system defined in classical and quantum mechanics
 by respectively $\{J_{j},J_{k}\}=\varepsilon_{jks}J_{s} $ and
 $[J_{j},J_{k}]=i\hbar\varepsilon_{jks}J_{s}$, $j,k=1,2,3$. When
 $J^2=J_1^2+J_1^2+J_3^2$ is fixed, the classical mechanics is
 confined within a sphere ${\cal S}_J$. Restricting the
 $su(2)$ Poisson structure to ${\cal S}_J$ yields a
 symplectic two form
 $\omega_2=J\sin\theta d\phi \wedge d\theta $, where
 $(\theta,\phi)$ is the conventional sphere coordinate. In quantum
 mechanics, $J^2=j(j+1)\hbar^2$, $j=\frac{1}{2},1,\frac{3}{2}...$.
 An eigenspace of $J^2$ is associated with a classical sphere
 ${\cal S}_J$, in which we shall assume $J=(j+\frac{1}{2})\hbar$ so
 that its phase space area (integral of $\omega_2$ on ${\cal S}_J$
 ) in unit $2\pi\hbar$ is $2j+1$, which directly corresponds to the
 dimension of the eigenspace. In our treatment of non-smooth
 systems, a prerequisite is that the phase space is the direct
 product of coordinate and momentum spaces. To meet this
 requirement, we write $(J\cos\theta+p_0,\phi)=(p,x)$, in which
 $\omega_2=d p\wedge d x$, and regard $(x,p)$ as the natural
 coordinate of the phase space of a mechanic system on a circle.
 Moreover, to ensure the right spectrum of $J_3=p-p_0$, we choose
 $p_0=0$ (or $\frac{1}{2}\hbar$) in the case of $j$ is an integer
 (or half integer). By this transformation in classical mechanics,
 we can treat the non-smoothness-enhanced tunneling in some spin
 systems.

{\sl Example 3.3} $$
  H(J_1,J_2,J_3)=\left \{ \begin{array}{lc}
   J_1^2-J_2^2+J_3^2 ~~~~&J_3{\ge}0,\\
   J_1^2-J_2^2 &J_3<0 .\\
 \end{array}
 \right.
 $$
 The corresponding classical system on a circle is
 $$
 H(x,p)=\left \{ \begin{array}{ll}
 [J^2 -(p-p_0)^2]\cos 2x+(p-p_0)^2 ~~~~& p{\ge}p_0,\\

 [J^2-(p-p_0)^2]\cos 2x            & p<p_0. \\
 \end{array}
 \right.
 $$
 From phase space portrait we know
 that energy levels in $(-J^2,0)$
 consist of 2-fold ND and according to Eq. (3.6-7),
 $$
 {\cal A}^{(0)}=\frac{-\hbar^2}{J^2
 \sin 2x_c}[W_3(2\pi-2x_c,\frac{p_0}{\hbar})+
 W_3(2x_c,\frac{p_0}{\hbar})e^{2i(\phi-\frac{\pi}{2})} +
 2W_3(\pi,\frac{p_0}{\hbar})e^{i(\phi-\frac{\pi}{2})}],
 $$
 where
 $x_c=\frac{1}{2}\cos^{-1}\xi$ with $\xi\equiv \epsilon/J^2$ and
 $\phi=\pi J(1-\sin x_c)/\hbar=\pi(j+\frac{1}{2})
 [1-(\frac{1-\xi}{2})^{\frac{1}{2}}]$.
 When $j$ is an integer,
 $$
|{\cal A}^{(0)}|=\frac {|\cos \phi|}{2(j+\frac{1}{2})^2(1-\xi)^2}
 $$
and when $j$ is a half integer, $$ |{\cal A}^{(0)}|
=\frac{1}{4(j+\frac{1}{2})^2(1-\xi^2)^{\frac{1}{2}}}
|\frac{3+\xi}{\sqrt{2}(1-\xi)^{\frac{3}{2}}}\sin\phi+\frac{1}{2}|
$$ (In this case, $\frac{1}{j+\frac{1}{2}}$ can be regarded
as an effective $\hbar$.) These relations give a good
description of the energy splitting when $j \gg 1$ (Fig.
6).

\section {Discussion}
 When the non-smooth system is controlled by a parameter
 $\lambda$, e.g. $V(x)\rightarrow \lambda V(x)$, it is
 easy to obtain a zero of ${\cal A}^{(0)}$ when $\lambda$ is
 continuously varied. One can naturally ask that
 whether the zero of ${\cal A}^{(0)}$
 predicts an exact degeneracy of energy level or
 it merely corresponds to a minimum of $\Delta \epsilon$.
 The answer turns out to be dependent upon the
 symmetry of the system. If the eigenstates
 involved in ND can be distinguished by
 different symmetries irrespective of the
 parameter, the energy difference between
 the two eigenstates should be a smooth
 function of $\lambda$,
 which is approximately given by
 $\frac{2\hbar}{T}{\cal A}^{(0)}$ or similar
 expression.
 In this case, the zero of ${\cal A}^{(0)}$
 indicates a nearby exact degeneracy.
 Of course, because of  the symmetry
 of $H$, this conclusion cannot be regarded as a
 violation of the well-known theorem of
 von Neumann and Wigner,
 which states that generically we must
 vary two parameters to create a
 degeneracy\cite{Wigner}.
 On the other hand, if the eigenstates
 cannot be restricted within different
 parameter-independent invariant subspaces,
 e.g.,when $H=p^2/2+\cos x +\lambda |\sin x|$,
 the zero of ${\cal A}^{(0)}$ generally corresponds
 to a minimum of $\Delta \epsilon$ where we
 must take the higher order
 corrections into account.

{\large ACKNOWLEDGEMENT}\\
 The author is grateful to the referees
 for pointing out Berry's related work
 to him and providing many suggestions
 to improve the manuscript.
 The author thanks Dr. W. M. Zheng
 for useful discussions.

\newpage
\parindent -5mm

\begin{center}
Figure Captions
\end{center}

Fig.1 Splitting of nearly degenerate energy levels. (a)-(d)
  for $H=H_1$, $H_2$, $H_3$ and $H_4$
 respectively.
 The numerical result of $\Delta \epsilon$
 (open circles),the spacing of semi-classical levels (dotted lines)
 and the semi-classical approximation of $\Delta \epsilon$ (solid lines)
 are shown at $\hbar=0.02$.
 The insets show the degenerate tori
 (solid line) in phase-space where $H$ is not smooth on the
 dotted lines.

Fig.2 Scaled energy splitting $\eta$ (open circles) and
$\eta^{(0)}$ (connected solid dots) in example 2.1 at $k=1$
to $4$ and $\hbar=0.05$.

Fig.3 Energy splitting $\Delta\epsilon$ (open circles) and
$\Delta\epsilon^{(0)}$ (solid lines) in example 2.2 at
$k=1$ to $4$ and $\hbar=0.04$.

Fig.4 Schematic figure show transition paths
$\gamma_1$ ($A\rightarrow A'$),
$\gamma_2$ ($B\rightarrow B'$)  and
closed path
$\gamma_{21}$ ($B\rightarrow B'\rightarrow A'\rightarrow A'\rightarrow
B')$.
 EBK quantization rule guarantees that
 $\phi_2-\phi_1$ (mod $2\pi$) is independent on the
 choice of real paths $B'\rightarrow A'$ on
 $O_{\epsilon}^{-}$ and $A\rightarrow B$ on
 $O_{\epsilon}^{+}$.

 Fig.5 $\eta$ (open circles) and $\eta^{(0)}$ (solid lines) in
 example 3.1 at $\hbar=0.02$. The inset shows three
 types of tori in phase space.
 The tori encircling point $(0,0)$ produce a
 semi-classically non-degenerate component of
 energy spectrum at $1<\epsilon\leq 2$, which has
 been excluded according to semi-classical
 criterion that the expectation value of $p^2$
 at the corresponding eigenstates is less than unity.

 Fig.6 $\eta$ (open circles) and $\eta^{(0)}$ (connected dots ) in
 example 3.3 at (a) $j=100$ and (b)$j=99\frac{1}{2}$.

\newpage

\parindent 2mm

{\bf \Large Appendix: Semi-classical Calculation of Energy Splitting}

We first consider the conventional Hamiltonian
$H=\frac{1}{2}p^2+V(x)$. Direct calculation show that $$
(-\frac{\hbar^2}{2}\frac{d^2}{dx^2}+V(x))\Psi^{\pm}(x)= (\epsilon+
Q(x))\Psi^{\pm}(x), \eqno(A.1) $$ with
$Q=-\frac{\hbar^2}{2}p^{1/2}(p^{-1/2})^{\prime \prime}$, where the
prime denotes derivation with respect to $x$ at fixed $\epsilon$.
Because $<\Psi_+|\Psi_+>=<\Psi_-|\Psi_->=1$ and $<\Psi_-|\Psi_+>\sim 0$,
the energy splitting calculated in the space spanned by $\Psi^{+}$
and $\Psi^{-}$ is given by $$ \Delta \epsilon=
2|<\Psi^-|Q|\Psi^+>|=
\frac{\hbar^2}{2T}|\int_0^{2\pi}[\frac{V^{\prime \prime}}{p^3}+
\frac{5(V^{\prime})^2}{2p^5}]\exp(i2s(x)/\hbar)dx|. \eqno(A.2)
$$

 Before evaluating $\Delta \epsilon$ according to Eq. (A.2),
 it is helpful to recall an useful mathematical result on
 asymptotic behavior of the Fourier coefficients of a
 non-smooth function. Let $f(x)$ be a sufficiently regular
 $2\pi$-periodic function on $R$. How its Fourier
 coefficients, defined by
 $$
 \widehat{f}(n)=\int_0^{2\pi}f(x)\exp(inx)dx, ~~~~~n\in Z,
 \eqno (A.3)
 $$
 decay
 when $n\rightarrow \pm\infty$ is
 basically determined by the analytic property of $f(x)$. If
 it is smooth, then $\widehat{f}(n)$ for large $n$ will
 approach zero faster than any power of $|n|^{-1}$, i.e.,
 $\lim_{|n|\rightarrow \infty} \widehat{f}(n)|n|^{\alpha}=0$
 for arbitrary $\alpha>0$ . On the other hand, if $f(x)$ is
 not smooth, the decay of $\widehat{f}(n)$ may follow a
 power law. In the simple case when $f(x)$ is the union of
 $N$ smooth segments on intervals
 $[x_i^{\ast},x_{i+1}^{\ast}]$,
 $x_1<x_2...<x_{N+1}=x_1+2\pi$, $\widehat{f}(n)$ can be
 expressed by asymptotic series
 $$
 \widehat{f}(n)=\sum_{l=0}^{\infty}\frac{i^{l+1}}{n^{l+1}}
 \sum_{j=1}^{N}\exp(inx_{j}^{\ast}){\bigwedge}_x^{l}f(x_{j}^{\ast}).
 \eqno(A.4) $$

 Let $s(x)=n\hbar\theta(x)$, we rewrite Eq. (A.2) as
 $$
 \Delta \epsilon=
 \frac{\hbar^2}{2T}|\int_0^{2\pi}[\frac{V^{\prime \prime}}{p^3}+
 \frac{5(V^{\prime})^2}{2p^5}]\frac{n\hbar}{p}\exp(i2n\theta)d\theta|.
 \eqno(A.5)
 $$
 Noticing the integrand apart from $\exp(i2n\theta)$ is
 unchanged in the semi-classical limit, according to Eq.
 (A.4), we have
 $$
 \Delta \epsilon=
 \frac{\hbar^{k+1}}{2^{k}T}|\sum_{j=1}^{N}\frac{\exp[2is(x_j^{\ast})/\hbar]}{p^{k+2}(x_j^{\ast})}
 {\bigwedge}_x^{k}V(x_j^{\ast})| +o(\hbar^{k+1}). \eqno(A.6)
 $$

 Then we consider Hamiltonian $H=E_{k}(p)+V(x)$. In order to
 evaluate energy splitting according to $\Delta
 \epsilon=2|<\Psi^{-}|H -\epsilon|\Psi^{+}>|$, it is
 instructive to go into some details about the momentum
 representation of $\Psi^{\pm}$. Write $$
 \Psi^{\pm}=\sum_{r=-\infty}^{\infty}\phi^{\pm}_{r}|r>,
 ~~~~~~~ <x|r>=\frac{1}{\sqrt{2\pi}}\exp(irx), \eqno(A.7) $$
 with $$ \phi^{\pm}_{r}=\frac{1}{\sqrt{2\pi}}\int_0^{2\pi}
 \Psi^{\pm}(x)\exp(-irx)dx. \eqno(A.8) $$ Since
 $\phi^{-}_{-r}=\phi^{+\ast}_{r}$, we shall focus on
 $\Psi^{+}$. The semi-classical limit of Eq. (A.8) should be
 calculated in two separate cases. In the classically
 permissible region (CPR), where $p(x)-r\hbar=0$ is
 satisfied by some $x\in [0,2\pi)$, stationary phase
 approximation can be adopted, which results $$
 \phi^{+}_{r}\approx\sqrt{\frac{\hbar}{T}}\sum_m
 \frac{1}{\sqrt{|V'(x_m)|}}
 \exp[i(s(x_m)/\hbar-rx_m-\sigma_m\pi/2)], \eqno(A.9) $$
 where$\{x_m\}$ are solutions of $p(x)-r\hbar=0$ and
 $\sigma_m={\rm sign}(V''(x_m))$. When $r\hbar$ is beyond
 CPR, by using expansion (A.4), we find
 $$
 \phi^{+}_{r}\approx \frac{(i\hbar)^{k+1}}{\sqrt{2\pi
 T}}\sum_{j=1}^{N}
 \frac{\exp[i(s(x_j^{\ast})/\hbar-rx_j^{\ast})]}
 {(p-r\hbar)^{k}{\frac{d}{dp}E_k}}
 \frac{d}{dp}[\frac{-1}{(p-r\hbar)\sqrt{\frac{d}{dp}E_k}}]|_{p=p(x_j^{\ast})}
 {\bigwedge}_x^{k}V(x_j^{\ast}).
 \eqno(A.10)
 $$
 From Eq. (A.9-10) we conclude that $\Psi^{+}$ consists of
 the main part distributed within CPR and two power-law-like long
 tails beyond CPR.
 ( As the non-smoothness of eigenfunction is resulted via the
 eigen equation
 from $V(x)$, this picture is also true for exact eigenfunction.)
 Furthermore, if the semi-classical momentum representation of
 $V\Psi^{+}$ is calculated in the similar procedure, one can find
 that the main part of $\Psi^{+}$ within CPR but its long
 tails approximately satisfies eigen equation
 $(E_k(p)+V(x))\Psi=\epsilon \Psi$, i.e.,
 $$
 \sum_{m=-\infty}^{\infty}(E_k(r\hbar)\delta_{m,0}+
 V_{m})\phi_{r+m}^{+}  \approx \epsilon\phi_{r}^{+},
 \eqno (A.11)
 $$
when $r\hbar \in {\rm CPR}$, where
 $$
 V_m=<0|V|m>\approx \frac{i^{k+1}}{2m^{k+1}\pi}
 \sum_{j=1}^{N}\exp(imx_{j}^{\ast}){\bigwedge}_x^{k}V(x_{j}^{\ast})
 ~~~(|m|\rightarrow \infty).
 \eqno (A.12)
 $$

Based on the above discussion, we know that
$$
\begin{array}{rl}
 <\Psi^{-}|E_k(p)|\Psi^{+}>&
 \displaystyle =\sum_{r=-\infty}^{\infty} \phi^{+}_{-r}E_k(r\hbar)\phi^{+}_r
 \approx
 \sum_{|r\hbar| \in {\rm CPR}}\phi^{+}_{-r}E_k(r\hbar)\phi^{+}_r\\\\
 &\displaystyle
 \approx
 (\sum_{\nu\hbar\in {\rm CPR}}\sum_{\mu=-\infty}^{\infty}+
 \sum_{-\mu\hbar\in {\rm CPR}}\sum_{\nu=-\infty}^{\infty})
 \phi^{+}_{-\nu}(\epsilon
 \delta_{\nu,\mu}-V_{\mu-\nu})\phi^{+}_{\mu}\\\\
 \end{array}
 \eqno(A.13) $$
 Compare the last expression with
 $$
 <\Psi^{-}|\epsilon-V|\Psi^{+}>= \sum_{\mu,\nu=-\infty}^{\infty}
 \phi^{+}_{-\nu}(\epsilon\delta_{\nu,\mu}-V_{\mu-\nu})\phi^{+}_{\mu}.
 \eqno(A.14)
 $$ The main contribution of Eq. (A.14) consists
 of three parts come from regions, (1)$\mu\hbar,\nu\hbar \in {\rm
 CPR}$, (2)$-\mu\hbar,-\nu\hbar \in {\rm CPR}$ and
 (3)$\mu\hbar,-\nu\hbar \in {\rm CPR}$ respectively.
 Eq. (A.13) contains only the former two parts while we
 can screen the last contribution by
 making a high frequency cut off of V(x), i.e., replacing it by $$
 V^{(0)}(x)=\sum_{|m|\leq k_c}V_{m}\exp(-i m x), \eqno (A.15) $$
 where $k_c$ is a large but fixed integer so that
 $V^{(1)}(x)=V(x)-V^{(0)}(x)$ is negligibly small. Therefore,
 $<\Psi^{-}|E_k(p)|\Psi^{+}>\approx
 <\Psi^{-}|\epsilon-V^{(0)}|\Psi^{+}>$, and consequently $$
 <\Psi^{-}|H-\epsilon|\Psi^{+}> \approx<\Psi^{-}|V^{(1)}|\Psi^{+}>=
 \frac{1}{T}\int_{0}^{2\pi}\frac{V^{(1)}} {\frac{d}{d p}
 E_{k}|_{p=p(x)}}\exp(2is(x)/\hbar)dx. \eqno (A.16) $$
 Observing that $V^{(1)}(x)\approx 0$
 and $\bigwedge_x^{j}V^{(1)}(x)=\bigwedge_x^{j}V(x)$
 for arbitrary $x\in [0,2\pi)$ and $j\geq 0$,
 by partial integrating Eq. (A.16) for successive $k+1$ times
 we obtain
 $$
<\Psi^{-}|H-\epsilon|\Psi^{+}>
=\frac{(i\hbar)^{k+1}}{2^{k+1}T}\sum_{j=1}^{N}\frac{\exp(2is(x_j^{\ast})/\hbar)}
{p^{k+1}\frac{d}{d p} E_{k}|_{p=p({x}_{j}^{\ast})}}
{\bigwedge}_x^{k}V({x}_{j}^{\ast})+o(\hbar^{k+1}), \eqno (A.17)
$$
which immediately leads to Eq. (2.4).

 Finally, it is worth pointing out that although the exact
 eigenstates have power-law tails beyond CPR, the leading
 term of $\Delta \epsilon$ actually does not relies on this
 detail. In fact, Eq. (A.16) essentially equals to $$
 \sum_{\mu\hbar, -\nu\hbar \in {\rm CPR}}
 \phi^{+}_{-\nu}V_{\mu-\nu}\phi^{+}_{\mu}, $$
 which is in nature controlled by the power-law decay
 of $\{V_m\}$ but $\{\phi_r^{+}\}$.
 Therefore, Eq. (A.16) can be reproduced
 from the highly localized semi-classical eigenfunctions
 corresponding to smoothed Hamiltonian
 $H^{(0)}=E_k(p)+V^{(0)}(x)$.

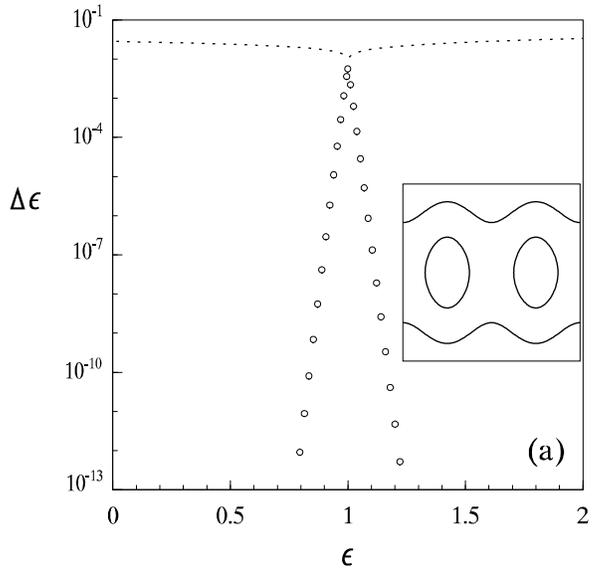
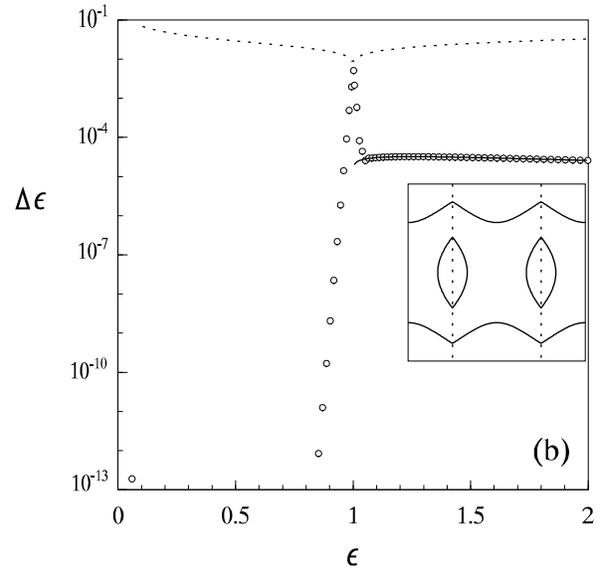
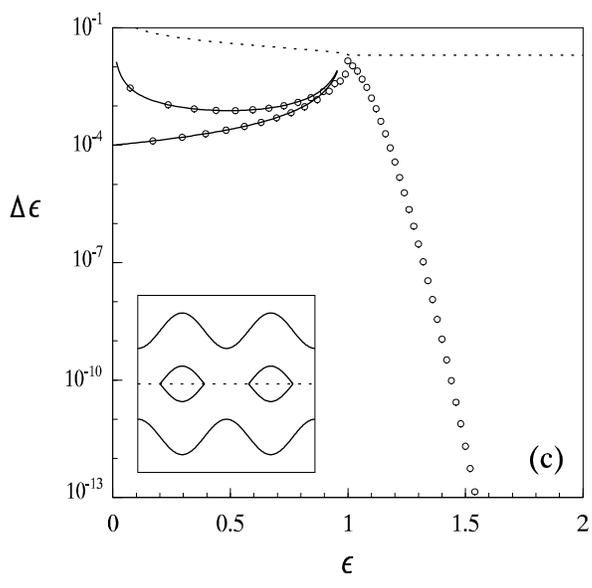
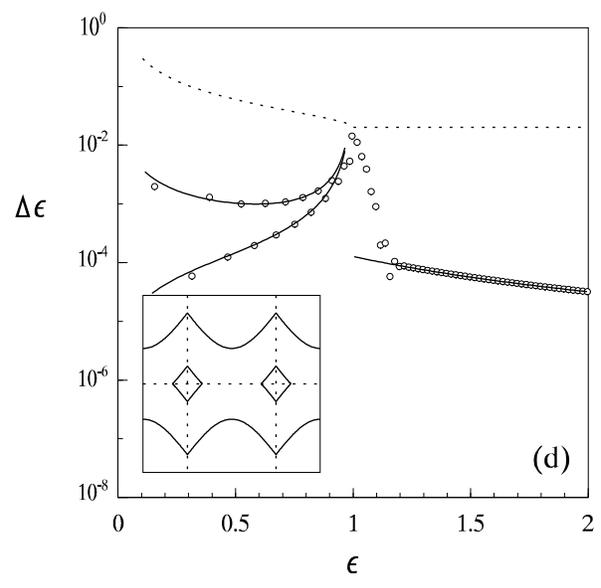
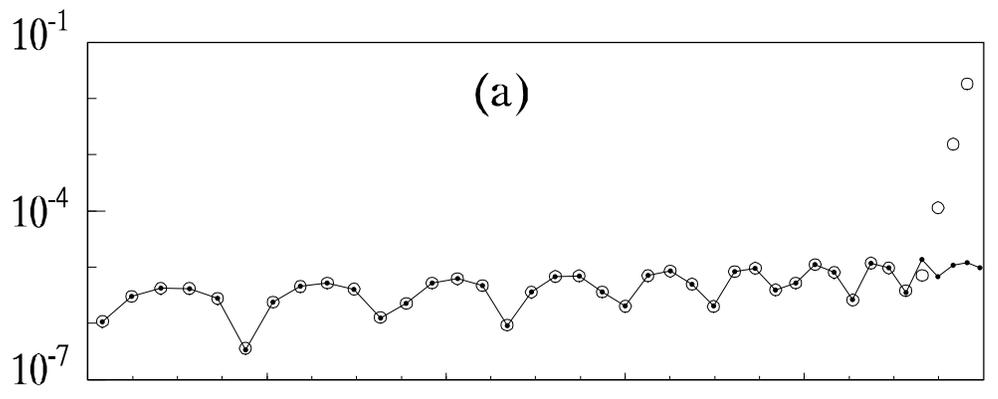
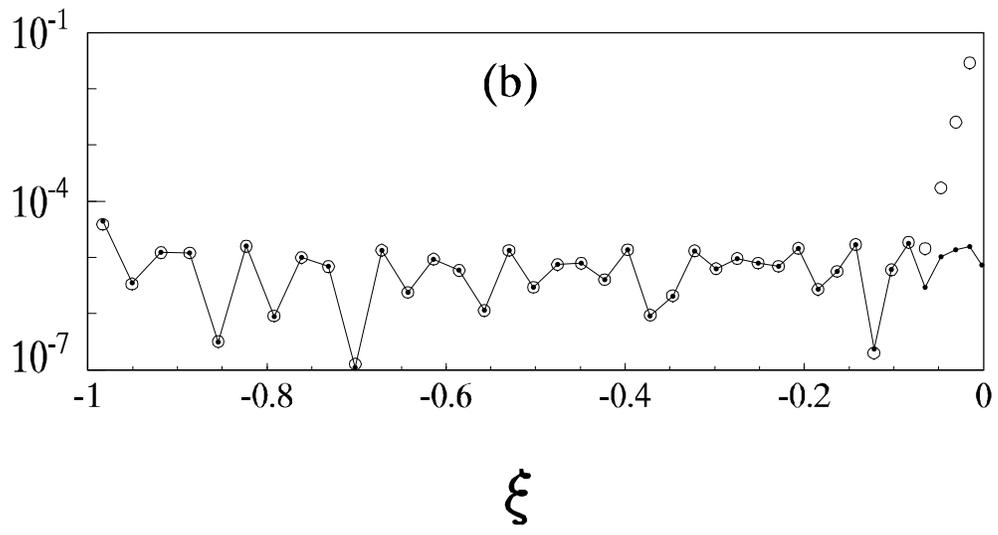
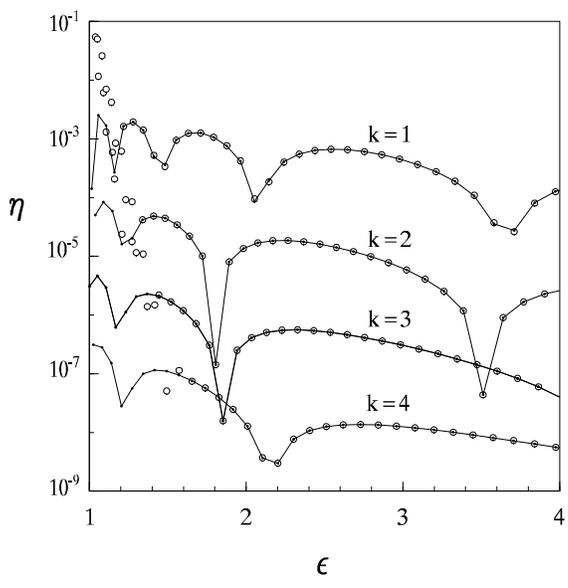
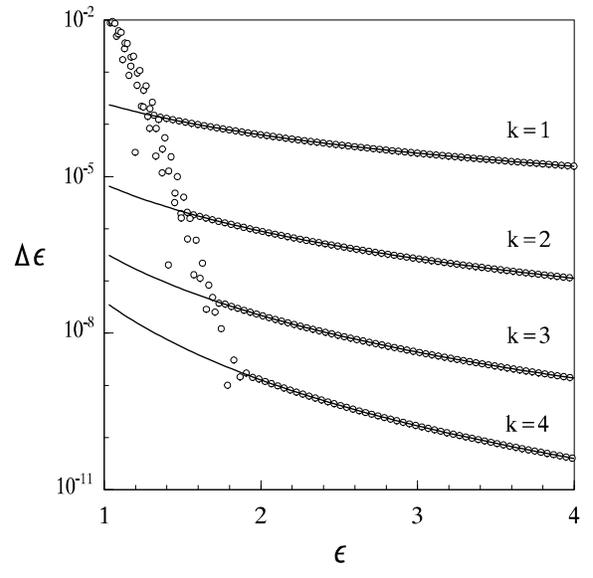
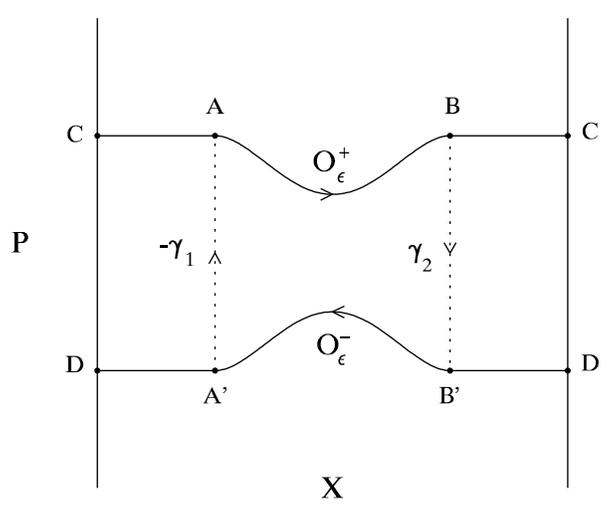
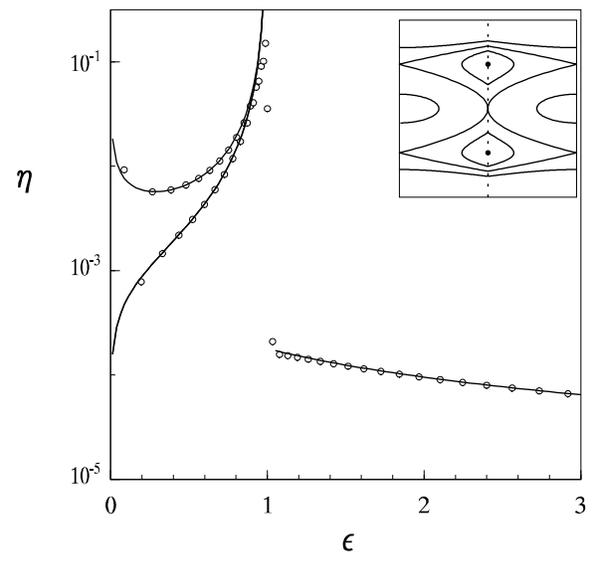
\end{document}